\documentclass{Interspeech}

\interspeechcameraready

\title{ClapFM-EVC: High-Fidelity and Flexible Emotional Voice Conversion with Dual Control from Natural Language and Speech}
\author[affiliation={1}]{Yu}{Pan}
\author[affiliation={3}]{Yanni}{Hu}
\author[affiliation={3}]{Yuguang}{Yang}
\author[affiliation={3}]{Jixun}{Yao}
\author[affiliation={3}]{Jianhao}{Ye}
\author[affiliation={3}]{Hongbin}{Zhou}
\author[affiliation={2}]{\\Lei}{Ma}
\author[affiliation={1}]{Jianjun}{Zhao}

\affiliation{Department of Information Science and Technology}{Kyushu University}{Japan}
\affiliation{Department of Computer Science}{The University of Tokyo}{Japan}
\affiliation{EverestAI}{Ximalaya Inc.}{China}
\email{panyu.ztj@gmail.com, ma.lei@acm.org, zhao@ait.kyushu-u.ac.jp}

\usepackage{comment}

\begin{document}

\maketitle

\begin{abstract}
Despite great advances, achieving high-fidelity emotional voice conversion (EVC) with flexible and interpretable control remains challenging. This paper introduces \textbf{\textit{ClapFM-EVC}}, a novel EVC framework capable of generating high-quality converted speech driven by natural language prompts or reference speech with adjustable emotion intensity. We first propose EVC-CLAP, an emotional contrastive language-audio pre-training model, guided by natural language prompts and categorical labels, to extract and align fine-grained emotional elements across speech and text modalities. Then, a FuEncoder with an adaptive intensity gate is presented to seamless fuse emotional features with Phonetic PosteriorGrams from a pre-trained ASR model. To further improve emotion expressiveness and speech naturalness, we propose a flow matching model conditioned on these captured features to reconstruct Mel-spectrogram of source speech. Subjective and objective evaluations validate the effectiveness of ClapFM-EVC.
\end{abstract}

\section{Introduction}

Emotional voice conversion (EVC) aims to convert the emotional state of source speech to a target category while preserving original content and speaker identity \cite{zhou2022mixed}.
Recently, EVC has garnered great attention within the speech processing realms and holds great potential for many practical applications such as voice assistant, audiobook production, and dubbing  \cite{zheng2020pre,chen2024takin,zhang2024speaker}.

In general, the key challenges for EVC lie in the accurate and efficient extraction, decoupling, and use of various speech attributes, such as the emotion \cite{pan2023msac,pan2024gmp}, content \cite{gulati2020conformer,kim2022squeezeformer}, and timbre \cite{wang2023cam,yao2024promptvc} information from speech signals.
With rapid progress in deep learning, existing EVC methods are generally based on generative adversarial networks (GANs) \cite{rizos2020stargan,zhou2020transforming,shankar2020non} and autoencoder models \cite{kim2020emotional,lu2022one,chen2023attention}.
StarGAN \cite{rizos2020stargan} used a cycle-consistent and class-conditional GAN to achieve EVC. \cite{chen2023attention} proposed AINN, an attention-based interactive disentangling network with a two-stage pipeline for fine-grained EVC.
However, the converted speech of these systems lacks emotional diversity, which is crucial for realistic speech synthesis \cite{qi2024towards}. To this end, several studies \cite{chen2023attention,qi2024towards,zhou2022emotion} shifted towards incorporating intensity control modules into EVC framework to allow more precise manipulation of emotional expression. 
Emovox \cite{zhou2022emotion} disentangled speaker style and controlled emotional intensity by encoding emotion in a continuous space.
EINet \cite{qi2024towards} predicted emotional class and intensity via an emotion evaluator and intensity mapper, incorporating controllable emotional intensity to enhance naturalness and diversity of emotion conversion.

Despite impressive advances, these approaches still face challenges. First, current EVC systems based on GANs and autoencoders, while promising, have great potential for improvements in emotional diversity, naturalness, and speech quality \cite{qi2024towards,chou2024toward}. Second, current methods typically rely on reference speech or categorical text labels as conditions to control a limited set of emotional expressions. Nevertheless, this paradigm not only imposes constraints on the user experience, but restricts the diversity of emotional expressions, while falling short in intuitiveness and interpretability of conveyed emotions.

To mitigate the aforementioned issues, this paper presents \textbf{ClapFM-EVC}, an innovative any-to-one EVC framework that enables flexible and intuitive control of emotion conversion in a user-friendly manner. To elaborate, we first propose EVC-CLAP (Contrastive Language Audio Pretraining), which is guided by both natural language prompts and emotional categorical labels, so as to extract and align the emotion features across speech-text modalities. 
Additionally, we introduce an end-to-end voice conversion (VC) model, termed AdaFM-VC, composed of pre-trained ASR model, FuEncoder and conditional flow matching (CFM) model. Using FuEncoder with an adaptive intensity gate (AIG), AdaFM-VC is able to integrate the captured emotional representations with Phonetic Posteriorgrams (PPGs) from a pre-trained ASR model HybridFormer \cite{yang2023hybridformer}, while allowing flexible control over the emotional intensity of the converted waveform. To further enhance naturalness and speech quality, we incorporate a CFM-based decoder \cite{yao2024stablevc,pan2024ctefm} that samples the output of the FuEncoder from random Gaussian noise and reconstructs the Mel-spectrogram of the source speech. 
During inference, EVC-CLAP can generate target emotional embeddings based on the given natural language prompt, and then AdaFM-VC leverages the target emotion vectors, source PPGs, and predefined emotional intensity to reconstruct the target Mel-spectrogram, which is ultimately converted into the target speech by a pre-trained vocoder \cite{lee2022bigvgan}. 
Extensive experiments and ablation studies demonstrate that our ClapFM-EVC significantly outperforms several existing EVC approaches in terms of emotional expressiveness, speech naturalness, and speech quality.

\begin{figure*}[htbp]
\centering
    \includegraphics[height=7.9cm,width=!]{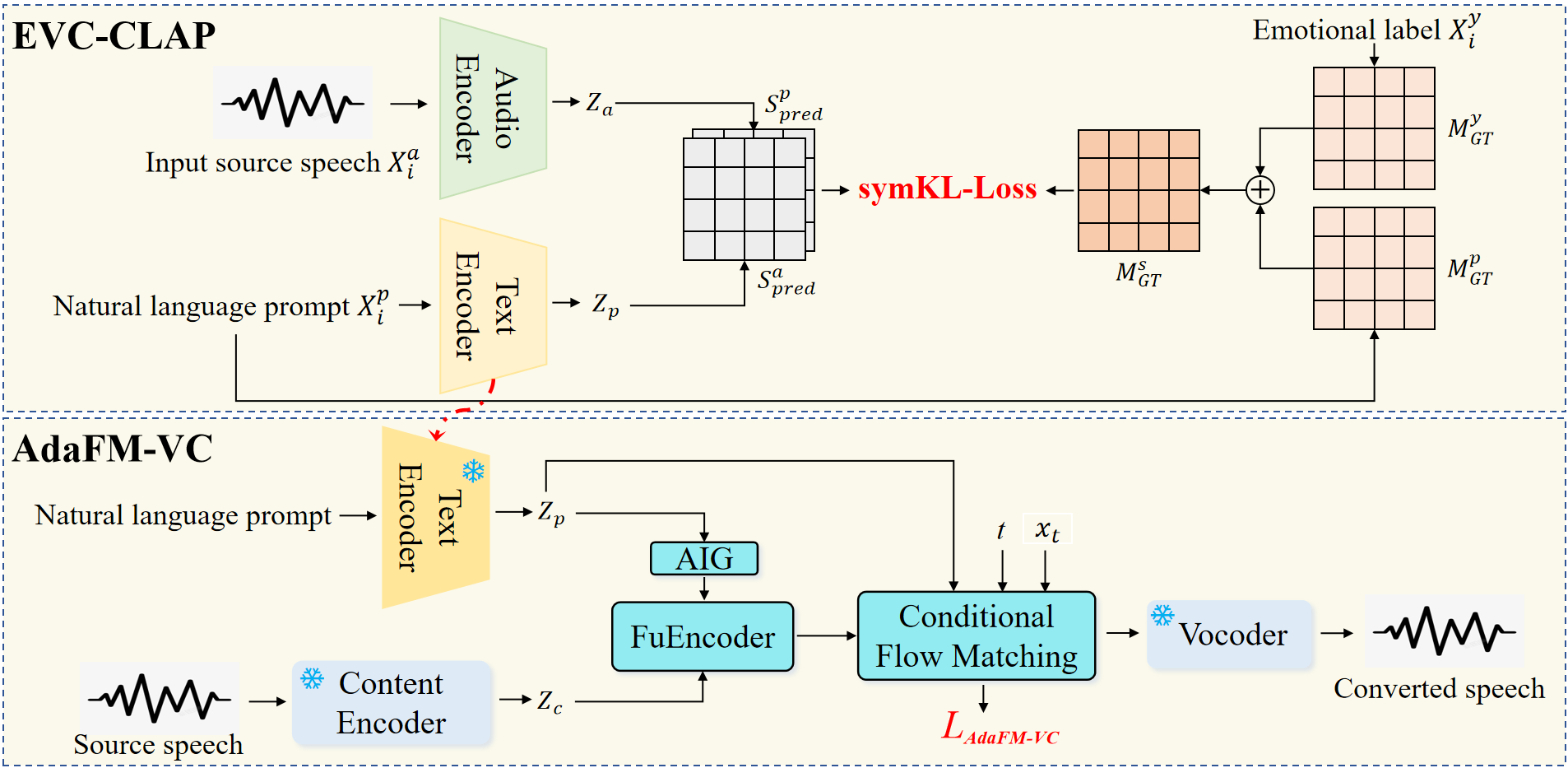}
    \caption{Overall training architecture of the proposed ClapFM-EVC framework.}
    \label{fig:clapfmevc}
\end{figure*}

\section{METHODOLOGY}
\label{sec:METHODOLOGY}
\subsection{System Overview}

As illustrated in Fig. \ref{fig:clapfmevc}, ClapFM-EVC can be characterized as a conditional latent model, where the proposed EVC-CLAP, FuEncoder, CFM-based decoder, as well as pretrained ASR \cite{yang2023hybridformer} and vocoder \cite{lee2022bigvgan} models serve as its core components.

Similarly to \cite{elizalde2023clap,pan2024gemo}, we first train EVC-CLAP using a symmetric Kullback-Leibler divergence based contrastive loss (symKL-loss) along with soft labels derived from natural language prompts and their corresponding categorical emotion labels. In this way, EVC-CLAP can effectively extract and align emotional representations across audio and text modalities, while enabling ClapFM-EVC to capture fine-grained emotional information conveyed by natural language prompts. 
Then, we train the AdaFM-VC using the obtained emotional elements and content representations extracted by EVC-CLAP and pre-trained HybridFormer, respectively.
The FuEncoder within AdaFM-VC facilitates the seamless integration of emotional and content characteristics, with its AIG module explicitly controlling the intensity of emotional conversion. 
Concurrently, the CFM model in AdaFM-VC samples the outputs of the FuEncoder from random Gaussian noise, and, conditioned on the target emotional vector produced by EVC-CLAP, it generates the Mel-spectrogram features of target speech. 
Finally, the generated Mel-spectrogram features are fed into a pre-trained vocoder to synthesize the converted speech. 

During inference, it is worth noting that our ClapFM-EVC framework provides three modes for obtaining the target emotional embeddings: (1) directly based on the provided reference speech; (2) directly based on the given natural language emotional prompt; and (3) EVC-CLAP retrieves relevant data from a pre-constructed high-quality reference speech corpus using specified natural language emotional prompt, subsequently extracting the target emotion elements from the retrieved speech.

\subsection{Soft-Labels-Guided EVC-CLAP}

Overall, the purpose of Emo-CLAP training is to minimize the distance between data pairs within the same class, while simultaneously maximizing the distance between pairs of data from different categories.

Assume that the input data pair is $\{X_i^a, X_i^{y}, X_i^{p}\}$, where \(X_i^a\) is the source speech, \(X_i^{y}\) and \(X_i^{p}\) denote its corresponding emotional label and natural language prompt, $i\!\in\![0, N]$ and N is the batch size.
Our EVC-CLAP first adopts a pre-trained HuBERT$^1$ \cite{hsu2021hubert} based audio encoder and a pre-trained XLM-RoBERTa$^2$ \cite{conneau2019unsupervised} based text encoder to compress \(X_i^a\) and \(X_i^{p}\) into two latent variables \(Z_a \! \in \mathbb{R}^{N\times D}\) and \(Z_p \! \in\mathbb{R}^{N\times D}\), where \(D\) equals 512, representing the hidden state dimension. Following this, we compute their corresponding similarity matrices $S^{a}_{\textit{pred}}$ and $S^{p}_{\textit{pred}}$ as:
\begin{equation}
    \begin{split}
        S^{a}_{\textit{pred}} = \varepsilon_a \times (Z_a \cdot {Z_p}^T) \\
        S^{p}_{\textit{pred}} = \varepsilon_t \times (Z_p \cdot {Z_a}^T)
    \end{split}
\end{equation}
where $\varepsilon_a$ and $\varepsilon_t$ are two learnable hyper-parameters, with their values empirically initialized to 2.3.
Subsequently, we employ symKL-loss to train Emo-CLAP with the guidance of the soft labels \(M^{s}_{\textit{GT}}\in\mathbb{R}^{N\times N}\) derived from \(X_i^{y}\) and \(X_i^{p}\). 
\begin{equation}
    \begin{split}
        M^{s}_{\textit{GT}} = \alpha_e M^{y}_{\textit{GT}} + (1-\alpha_e) M^{p}_{\textit{GT}}
    \end{split}
\end{equation}
where $\alpha_e$ is a hyper-parameter to adjust $M^{y}_{\textit{GT}}$ and $M^{p}_{\textit{GT}}$, empirically set to 0.2 in our case. 
In detail, if the categorical emotional labels or natural language prompt labels of different data pairs within the same batch are identical, their corresponding ground truth is assigned a value of 1; otherwise, it is set to 0. To ensure the consistency of the label distributions across the batch, the class similarity matrices \(M^{E}_{\textit{GT}}\) and \(M^{P}_{\textit{GT}}\) are normalized such that the sum of each row equals 1, effectively capturing the relative similarity between data pairs.
Therefore, the training loss of EVC-CLAP can be formulated as: 
\begin{equation}
    \begin{split}
        L_{\text{\textit{symKL}}} = \frac{1}{4} \Bigg( 
        \text{\textit{KL}}\left( S^{a}_{\textit{pred}}|| M^{s}_{\textit{GT}} \right) + 
        \text{\textit{KL}}\left( \tilde{M^{s}_{\textit{GT}}}|| S^{a}_{\textit{pred}} \right) \\
        + \text{\textit{KL}}\left( S^{p}_{\textit{pred}}|| M^{s}_{\textit{GT}} \right) + 
        \text{\textit{KL}}\left( \tilde{M^{s}_{\textit{GT}}}|| S^{p}_{\textit{pred}} \right)
        \Bigg)
    \end{split}
\end{equation}
\begin{equation}
    \tilde{M^{s}_{\textit{GT}}} = (1 - \alpha) \cdot M^{s}_{\textit{GT}} + \frac{\alpha}{N}
\end{equation}
\begin{equation}
    \begin{split}
        \text{\textit{KL}}(S|| M) = \sum_{i,j} S(i,j) \log \frac{S(i,j)}{M(i,j)}
    \end{split}
\end{equation}
where $\alpha$ is a hyper-parameter, empirically set to $1 \!\times\! 10^{-8}$.

{
\let\thefootnote\relax
\footnote{$^1$https://huggingface.co/TencentGameMate/chinese-hubert-large
}
\footnote{$^2$https://huggingface.co/FacebookAI/xlm-roberta-base}
}

\subsection{AdaFM-VC}

\subsubsection{FuEncoder with AIG}

As a pivotal intermediate component within ClapFM-EVC, FuEncoder aims to seamlessly integrate content features extracted by HybridFormer with emotional embeddings derived from EVC-CLAP, while offering flexible control over the emotion intensity through the adaptive intensity gate, namely AIG.

Detailed, FuEncoder comprises a preprocessing network (PreNet), a positional encoding module, an AIG module, an adaptive fusion module, and a linear mapping layer. 
First, PreNet is presented to compress the source content features $Z_{c}$ to a latent space, preventing overfitting through a dropout mechanism. 
Next, a positional encoding module is advocated to employ sinusoidal positional encoding to extract the positional characteristics of \( Z_{c} \) and performs element-wise addition with \( Z_{c} \) to ensure that FuEncoder learns its sequential and structural information.
Afterwards, we propose an AIG module to multiply a learnable hyperparameter by the EVC-CLAP's emotional features to flexibly adjust the emotional intensity. As the core of FuEncoder, the adaptive fusion module consists of multiple fusion blocks, each of which contains a multi-head self-attention layer, two emotion adaptive layer norm layers \cite{min2021meta}, and a position-wise feed-forward network layer, enabling efficient fusion of content and emotional information, thus generating rich embedding representations that contain both linguistic and emotional characteristics. The fused features are ultimately mapped to the specific dimensions \(f \! \in \! R^{B \times T \times D} \) through a fully connected layer.

\subsubsection{Conditional Flow Matching-based Decoder}
To further enhance speech naturalness and speech quality, we incorporate an optimal transport (OT)-based CFM model to reconstruct the target Mel-spectrogram \( x_1 = p_1(x) \) from a standard Gaussian noise \( x_0 = p_0(x) = \mathcal{N}(x;0,I) \). 
To elaborate, conditioned on the captured EVC-CLAP's emotional embeddings, an OT flow \( \psi_{t}: [0, 1] \times \mathbb{R}^d \rightarrow \mathbb{R}^d \) is adopted to train our CFM-based decoder, which consists of 6 CFM blocks with timestep fusion. Each CFM block contains a ResNet \cite{he2016deep} module, a multi-head self-attention \cite{vaswani2017attention} module and a FiLM \cite{perez2018film} layer. By utilizing an ordinary differential equation to model a learnable and time-dependent vector field \( v_t: [0, 1] \times \mathbb{R}^d \rightarrow \mathbb{R}^d \), the flow can approximate the optimal transport path from \( p_0(x) \) to the target distribution \( p_1(x) \): 
\begin{equation}
    \label{eq:ode}
    \frac{d}{dt} \psi_t(x) = v_t(\psi_t(x), t)
\end{equation}
where \( \psi_0(x) = x \) and \( t \in [0, 1] \).
Besides, drawing inspiration from previous works \cite{yang2024takin}, which suggest adopting straighter trajectories, we simplify the OT flow formula as follows: 
\begin{equation}
    \label{eq:111}
    \psi_{t,z}(x) = \mu_t(z) + \sigma_t(z)x
\end{equation}
where \( \mu_t(z) = t z \), \( \sigma_t(z) = (1 - (1 - \sigma_{\text{min}}) t) \), and \( z \) represents the random conditioned input. \( \sigma_{\text{min}} \) denotes the minimum standard deviation of the white noise introduced to perturb individual samples, with its value empirically set to 0.0001.

Consequently, the training loss for AdaFM-VC is defined as:
\begin{equation}
    \mathcal{L} = \mathbb{E}_{t,p(x_0),q(x_1)}\Vert(x_1-(1-\sigma)x_0)-v_t({\psi_{t,x_1}}(x_0)|h)\Vert^2
\end{equation}
where \( x_0 \! \sim p(x_0) \), \( x_1 \!  \sim q(x_1) \), \( t \! \sim U[0,1] \), \( q(x_1) \) denotes the true yet potentially non-Gaussian distribution of the data, \( h \) refers to the conditional emotion embeddings extracted by EVC-CLAP.

\section{EXPERIMENTS}
\label{sec:EXPERIMENTS}

\subsection{Experimental Setups}

\subsubsection{Datasets}
Since no open-source EVC corpus with comprehensive emotional natural language prompts is currently available, we leverage an internally developed expressive single-speaker Mandarin corpus for training the proposed ClapFM-EVC system. This corpus encompasses 20 hours of speech data sampled at 24 kHz. From this, we specifically selected 12,000 utterances representing 7 original categorical emotion classes (neutral, happy, sad, angry, fear, surprise, disgust). To ensure high-quality annotations, we enlisted 15 professional annotators to provide natural language prompts for the selected waveforms.

\subsubsection{Implementation Details}

In all experiments, the proposed EVC-CLAP and AdaFM-VC models are trained end-to-end using 8 NVIDIA RTX 3090 GPUs. For training the EVC-CLAP model, the Adam optimizer is employed with an initial learning rate of $1 \times 10^{-5}$ and a batch size of 16. All models are trained for 40 epochs within the PyTorch framework. Afterwards, the AdaFM-VC approach is trained using the AdamW optimizer over 500,000 iterations on the same GPU setup, with an initial learning rate of $2 \times 10^{-4}$ and a batch size of 32. During inference, the Mel-spectrogram of the target waveform is sampled using 25 Euler steps within the CFM-based decoder, with a guidance scale set to 1.0.

% \subsubsection{Comparative Baseline Systems}
% We compare CTEFM-VC with four SOTA VC systems to evaluate its zero-shot VC performance. 
% The first baseline, DiffVC \cite{popov2021diffvc}, is a zero-shot VC approach employing diffusion models alongside a fast maximum likelihood sampling scheme. 
% The second, NS2VC, is a modified variant of NaturalSpeech2 \cite{shen2023naturalspeech} that integrates diffusion models and neural codecs. 
% The third, VALLE-VC \cite{wang2023valle}, is adapted to perform voice conversion by replacing the original phoneme input with semantic tokens extracted from a supervised model. 
% The final baseline, SEFVC \cite{li2024sef}, utilizes a position-agnostic cross-attention mechanism to incorporate the target timbre from reference speech. To ensure a fair comparison, all methods are trained on the same dataset.

\subsubsection{Evaluation Metric}

To assess speech quality and emotion similarity of ClapFM-EVC, we perform subjective and objective evaluations. 
Regarding speech quality, a variety of objective metrics are used, including Mel-cepstral distortion (MCD), root mean squared
 error (RMSE), character error rate (CER), and predicted MOS (UTMOS). For emotion similarity, we employ a pretrained speech emotion recognition model$^3$ \cite{ma2023emotion2vec} to compute the cosine similarity of emotion embeddings between the converted and reference speech, referred to as the emotion embedding cosine similarity (EECS).
The CER and UTMOS are calculated using pretrained CTC-based ASR$^4$ and MOS prediction$^5$ approaches.
Besides, the subjective Mean Opinion Score (MOS) with a 95\% confidence interval is used to measure naturalness (nMOS) and emotion similarity (eMOS). In practice, we invite 12 professional raters to participate in the evaluation. The scoring scale ranges from 1 to 5, with increments of 1, where higher scores indicate better performance.
The audio samples are available online$^6$.

{
\let\thefootnote\relax
\footnote{$^3$https://github.com/ddlBoJack/emotion2vec}
\footnote{$^4$https://huggingface.co/facebook/hubert-large-ls960-ft}
\footnote{$^5$https://github.com/tarepan/SpeechMOS}
\footnote{$^6$https://anonymous.4open.science/w/clapfm-evc-ACF1}
\footnote{$^7$https://github.com/glam-imperial/EmotionalConversionStarGAN}
\footnote{$^8$https://github.com/KunZhou9646/seq2seq-EVC}
\footnote{$^9$https://github.com/KunZhou9646/Mixed\_Emotions}
}

\begin{table*}[ht]
\centering
\caption{Overall comparison results of the speech quality and emotion similarity between our proposed ClapFM-EVC system and other SOTA baseline methods using reference speech. nMOS and eMOS are presented with 95\% confidence intervals.}
\label{tab:overall}
\begin{tabular}{cccccccc}
\hline
\textbf{Model} & MCD ($\downarrow$)  & RMSE ($\downarrow$)  & CER ($\downarrow$) & UTMOS ($\uparrow$) & nMOS ($\uparrow$) & EECS ($\uparrow$) & eMOS ($\uparrow$)  \\ \hline
StarGAN-EVC  \cite{rizos2020stargan}   & 8.85    & 19.48  & 13.07   & 1.45   &  2.09 $\pm$ 0.12    & 0.49 & 1.97 $\pm$ 0.09  \\
Seq2seq-EVC  \cite{zhou2021limited} & 6.93     & 15.79 & 10.56   & 1.81   &  2.52 $\pm$ 0.11    & 0.54 & 2.23 $\pm$ 0.11  \\
MixEmo  \cite{zhou2022speech} & 6.28    & 13.84  & 8.93   & 2.09   &  2.98 $\pm$ 0.07    & 0.65 & 2.58 $\pm$ 0.13  \\
ClapFM-EVC   & \textbf{5.83}   & \textbf{10.91}  & \textbf{6.76}  & \textbf{3.68}    & \textbf{4.09 $\pm$ 0.09}    & \textbf{0.82}  & \textbf{3.85 $\pm$ 0.06} \\ \hline
\end{tabular}
\end{table*}

\subsection{Main Results}

\subsubsection{EVC by Reference Speech}
To evaluate the performance of ClapFM-EVC, we compare it with several existing EVC methods, i.e., StarGAN-EVC$^7$  \cite{rizos2020stargan}, Seq2seq-EVC$^8$ \cite{zhou2021limited}, and MixEmo$^9$ \cite{zhou2022speech}. 
Since these baselines employ the reference waveform to facilitate EVC, we first examine their performance using reference speech. 

As evidenced in Table \ref{tab:overall}, our ClapFM-EVC consistently attains state-of-the-art performance in both speech quality and emotion similarity. 
With regard to emotion similarity, ClapFM-EVC exhibits significant advancements over baseline approaches, achieving notable relative improvements of at least 26.2\% in EECS and 53.1\% in eMOS, respectively.
This indicates that our proposed ClapFM-EVC framework has a remarkable capability to precisely capture and effectively transfer the target emotional characteristics during emotional voice conversion.
Regarding speech quality, the experimental results reveal that ClapFM-EVC exhibits superior results across multiple objective metrics. Detailed, ClapFM-EVC shows enhanced performance by achieving the lowest values in MCD, RMSE, and CER metrics, respectively. 
Moreover, subjective evaluation confirms that ClapFM-EVC gains the highest scores in both nMOS and UTMOS, with relative improvements of 37.2\% and 49.2\%, respectively, over the best-performing baseline method. These results underscore its exceptional capability in maintaining superior perceptual quality.

% As a consequence, these evaluations collectively highlight the effectiveness and robustness of our ClapFM-EVC approach.

\subsubsection{EVC by Natural Language Prompt}
To compare the performance of ClapFM-EVC when using reference speech (Reference) versus natural language prompts (Prompt), we further conduct an ABX preference test.

\begin{figure}[htbp]
\centering
    \includegraphics[height=2.3cm,width=!]{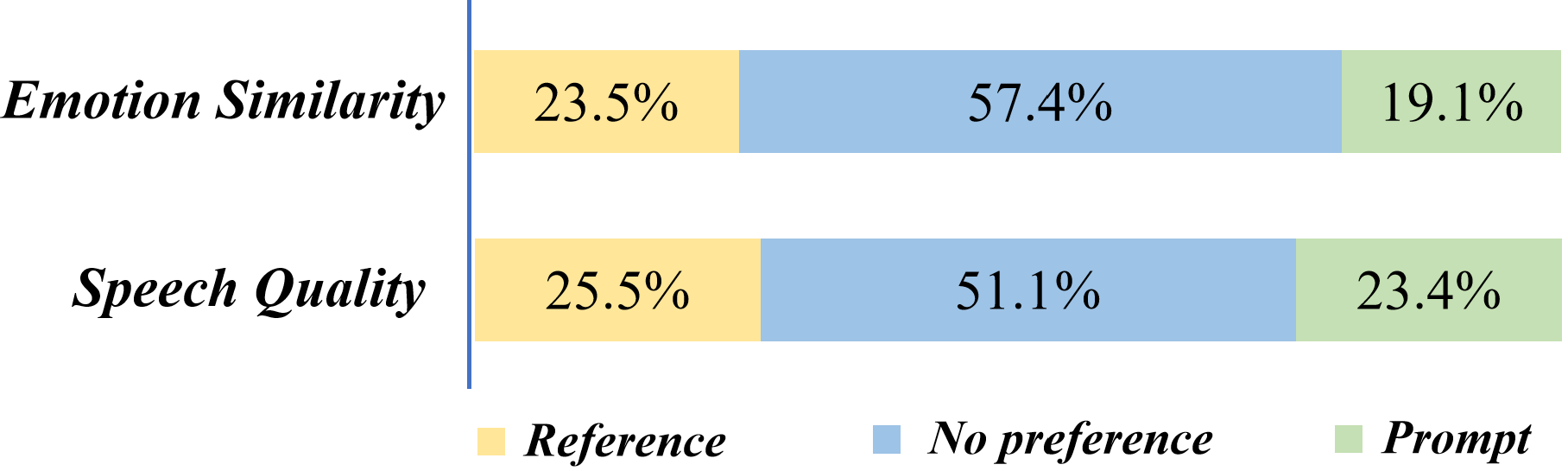}
    \caption{The ABX preference test results compare the Reference with Prompt.}
    \label{fig:abx}
\end{figure}

As shown in Fig. \ref{fig:abx}, the first test aims to evaluate the emotional similarity between the converted speech driven by \textit{Reference} and \textit{Prompt}. 47 participants were asked to rate speech samples generated by \textit{Prompt}, with \textit{Reference} serving as the benchmark, on a scale from -1 to 1. Here, -1 indicates that the converted speech driven by \textit{Reference} shows better emotion similarity, and 0 indicates no preference. The results revealed that 57.4\% of participants selected "no preference," while 19.1\% favored the "\textit{Prompt}," suggesting that ClapFM-EVC can effectively control the emotional expression of converted speech via \textit{Prompt}. In addition, we assess the quality of the converted speech relative to ground truth samples. Participants were required to choose the converted speech sample that is closer to ground truth in speech quality. As shown in the figure, the preference rates for speech driven by \textit{Reference} and \textit{Prompt} are 25.5\% and 23.4\%, showcasing that ClapFM-EVC is able to achieve high-quality EVC driven by \textit{Prompt}.

% {
% \let\thefootnote\relax
% \footnote{$^6$https://github.com/KunZhou9646/seq2seq-EVC}
% \footnote{$^7$https://github.com/KunZhou9646/Emovox}
% \footnote{$^8$https://github.com/KunZhou9646/Mixed\_Emotions}
% }

\subsection{Ablation Study}

To evaluate the contributions and validity of each component of the proposed system, 
we conduct ablation studies. All results are summarized in Table \ref{tab:ablation}.

\begin{table}[ht]
\centering
\caption{Ablation results of the proposed ClapFM-EVC driven by natural language prompts. 'w/o emo label' denotes removing emotional categorical labels when training EVC-CLAP, 'w/o symKL' represents replacing symKL-loss with KL-Loss, 'w/o AIG' denotes removing the AIG module of AdaFM-VC.}
\label{tab:ablation}
\renewcommand\arraystretch{1.2}
\resizebox{1.0\linewidth}{!}{
\begin{tabular}{ccccc}
\hline
\textbf{Model} & \textbf{UTMOS} ($\uparrow$) & \textbf{nMOS} ($\uparrow$) & \textbf{EECS} ($\uparrow$) & \textbf{eMOS} ($\uparrow$) \\ 
\hline
ClapFM-EVC   & \textbf{3.63}   & \textbf{4.01 $\pm$ 0.06}           & \textbf{0.79}           & \textbf{3.72 $\pm$ 0.08}   \\
\quad \quad w/o emo label   & 3.61           & 3.96 $\pm$ 0.11         & 0.66           & 3.01 $\pm$ 0.07   \\
\quad w/o symKL   & 3.57          & 3.89  $\pm$ 0.05       & 0.71           & 3.28 $\pm$ 0.08 \\
\quad w/o AIG      & 3.25  & 3.62 $\pm$ 0.12  & 0.74  & 3.52 $\pm$ 0.05   \\ \hline
\end{tabular}
}
\end{table}

From the table above, we can easily reach the following conclusions: (1) In the absence of categorical emotional labels, the EECS and eMOS scores exhibit a significant decline, while the speech quality metrics remain largely unaffected. This demonstrates the effectiveness and rationality of the proposed soft-label-guided training strategy in our EVC-CLAP. (2) When training EVC-CLAP with KL-Loss, the EECS and eMOS values showed a relative performance drop of 10.1\% and 11.8\%, indicating that the proposed symKL-Loss can effectively enhance the emotion representation capability of EVC-CLAP. (3) The removal of the AIG module leads to a notable deterioration in speech quality and a slight performance reduction in emotional similarity. This underscores the critical role of the proposed AIG module in adaptively integrating content and emotional characteristics, thereby improving the overall performance of our proposed ClapFM-EVC system.

\section{CONCLUSIONS}
\label{sec:CONCLUSIONS}
In this study, we propose ClapFM-EVC, an innovative and effective high-fidelity any-to-one EVC framework that features flexible and interpretable emotion control along with adjustable emotion intensity. 
Specifically, the proposed ClapFM-EVC initially employs EVC-CLAP to extract and align emotional elements across audio-text modalities. To enhance the emotional representational capacity, we utilize symKL-Loss to train the proposed EVC-CLAP model, guided by soft labels derived from the natural language prompts and their corresponding categorical emotion labels. 
To improve speech quality and naturalness, we subsequently introduce a flow matching-based AdaFM-VC model and a pretrained Vocoder to achieve high-fidelity emotional voice conversion.
Extensive experiments indicate that our proposed ClapFM-EVC is capable of generating converted speech with precise emotion control and high speech quality driven by natural language prompts.

\bibliographystyle{IEEEtran}
\bibliography{clapfmevc}

\end{document}